\documentclass[useAMS,usedcolumn,usenatbib,usegraphicx,sort]{mn2e}

\usepackage{graphicx}

\usepackage{epsfig,graphicx,natbib}
\usepackage{amssymb}
\usepackage{gensymb}
\usepackage{amsfonts}
\usepackage{amsmath}
\usepackage{color}
\usepackage{lscape,longtable}
\usepackage{times}

\long\def\symbolfootnote[#1]#2{\begingroup%
\def\thefootnote{\fnsymbol{footnote}}\footnote[#1]{#2}\endgroup} 

\title[Discovery of the First z $\boldmath \ge$ 6 Quasar From the Dark Energy Survey]
{DES J0454$-$4448: Discovery of the First Luminous z $\ge$ 6 Quasar from the 
Dark Energy Survey} 
\author[S.L. Reed, McMahon R.G., M. Banerji et al.]
{\parbox{\textwidth} 
{S. L. Reed$^{1,2}$\thanks{E-mail: sr525@ast.cam.ac.uk}, 
R. G. McMahon$^{1,2}$, M. ~Banerji$^{1,2,3}$, G. D. ~Becker$^{1,2,6}$, E.
Gonzalez-Solares$^{1}$, P. ~Martini$^{4,5}$, F. ~Ostrovski$^{1,2}$, M.
~Rauch$^7$, T.~Abbott$^9$, F.~B.~Abdalla$^3$, S.~Allam$^{10,6}$,
A.~Benoit-Levy$^{3}$, E.~Bertin$^{11}$, E.~Buckley-Geer$^{10}$,
D.~Burke$^{12}$, A.~Carnero Rosell$^{13}$, L.~N.~da Costa$^{18,13}$,
C.~D'Andrea$^{14}$, D.~L.~DePoy$^{15}$, S.~Desai$^{16}$, H.~T.~Diehl$^{10}$,
P.~Doel$^{3}$, C.~E Cunha$^{17}$, J.~Estrada$^{10}$, A.~E.~Evrard$^{8}$,
A.~Fausti Neto$^{18}$, D.~A.~Finley$^{10}$, P.~Fosalba$^{19}$,
J.~Frieman$^{10}$, D.~Gruen$^{16,20}$, K.~Honscheid$^{4}$, D.~James$^{9}$,
S.~Kent$^{10}$, K.~Kuehn$^{22}$, N.~Kuropatkin$^{10}$, O.~Lahav$^{3}$,
M.~A.~G.~Maia$^{18,13}$, M.~Makler$^{23}$, J.~Marshall$^{15}$,
K.~Merritt$^{10}$, R.~Miquel$^{24}$, J.~Mohr$^{16}$, B.~Nord$^{10}$, R.~Ogando$^{13}$,
A.~Plazas$^{25}$, K.~Romer$^{14}$, A.~Roodman$^{12,17}$, E.~Rykoff$^{12}$,
M.~Sako$^{26}$, E.~Sanchez$^{27}$, B.~Santiago$^{28}$, M.~Schubnell$^8$,
I.~Sevilla$^{27}$, C.~Smith$^{9}$, M.~Soares-Santos$^{10}$,
E.~Suchyta$^{4,21}$, M.~E.~C.~Swanson, G.~Tarle$^{8}$, D.~Thomas$^{14,30}$,
D.~Tucker$^{10}$, A.~Walker$^{9}$ and R.~H.~Wechsler$^{29,17}$}\\ 
Affliations at end of paper.
}

\begin{document}
\date{Accepted by MNRAS 2015 September 1. Received 2015 August 27; in original
form 2015 April 13\\ Accepted version. This is a pre-copyedited,
author-produced version of an article accepted for publication in MNRAS
following peer review.}
\maketitle

\begin{abstract}
We present the first results of a survey for high redshift, z $\ge$ 6,
quasars using izY multi-colour photometric observations from the Dark Energy
Survey (DES).  Here we report the discovery and spectroscopic confirmation of
the $\rm z_{AB}, Y_{AB}$ = 20.2, 20.2 (M$_{1450}$ = $-$26.5) quasar 
DES J0454$-$4448 with a redshift of z = 6.09$\pm$0.02 based on the onset of the
Lyman-$\alpha$ forest and a HI near zone size of 4.1$_{-1.2}^{+1.1}$ proper
Mpc.  The quasar was selected as an i-band drop out with i$-$z = 2.46 and
z$_{AB} < 21.5$ from an area of $\rm \sim$300 deg$^2$. It is the brightest of
our 
43 candidates and was identified for spectroscopic follow-up solely based on
the DES i$-$z and z$-$Y colours.  The quasar is detected by WISE and has
$W1_{AB} = 19.68$. The discovery of one spectroscopically confirmed quasar
with 5.7 $<$ z $<$ 6.5 and z$_{AB} \leq$ 20.2 is consistent with recent
determinations of the luminosity function at z $\sim$ 6. DES when completed
will have imaged $\rm \sim$5000 deg$^2$ to $Y_{AB}$ = 23.0 ($5\sigma$ point
source) and we expect to discover 50-100 new quasars with
z $>$ 6 including 3-10 with z $>$ 7 dramatically increasing the numbers of
quasars currently known that are suitable for detailed studies.

\end{abstract}

\begin{keywords} dark ages, reionization, first stars --- galaxies: active ---
galaxies: formation --- galaxies: high redshift --- quasars: individual:
DES~J0454-4448 \end{keywords}

\section{Introduction}

High redshift quasars act as important probes of the very early
Universe. They can be used to study the formation and evolution of the
first supermassive black holes and to provide insight into the
properties of the inter-galactic medium (IGM). The Ly$\alpha$ forest
and associated absorption lines blueward of the Ly$\alpha$ peak in a
high-redshift quasar spectrum give a direct measurement of the neutral
hydrogen in the IGM \citep{Fan2006b}. Detailed studies of the ionized near zone
in quasars can be used to investigate the ionization state and evolution of the
neutral hydrogen fraction at z$>$6 \citep{Bolton2007}.  Cosmologically
distributed intervening quasar absorption lines allow us to map the
distribution of gas in the IGM as well as the metal content.  Identifying
larger numbers of high redshift quasars would enable such studies to be
conducted further back in cosmic time along independent lines of sight through
the IGM.

Spectra of quasars from z $\sim$ 2 \citep{Gunn1965} up to z $\sim$
6 \citealt{Fan2006b, Becker2007} indicate that the IGM is highly ionised ($n_{\rm
{HI}}/n_{\rm H} \le 10^{-4}$) and therefore that reionization was complete at
earlier epochs. The 2$\sigma$ lower-bound for the Epoch of Reionization is
estimated to be at $z\sim6$ from Cosmic Microwave Background (CMB) measurements
\citep{Planck2015XIII}. The discovery of a large sample of new high redshift
quasars at $z>6$ would allow this important epoch to be studied in detail.
Even the discovery of a few new objects can be valuable if they have
extreme properties, such as being very bright (\citep{Wu2015}) or
very distant \citep{Mortlock2011}.

Previous surveys for high redshift ($z \ge 6.0$) quasars (e.g.
\citealt{Venemans2015, Carnall2015, Willott2010, Jiang2009, Mortlock2012,
Venemans2013, Fan2006b}) have already led to the discovery of $\sim$ 50 quasars
at $z \ge 6.0$.  Most of these surveys have employ purely optical photometric
datasets such as those provided by the Sloan Digital Sky Survey (SDSS) and the
Canada France Hawaii Telescope Survey (CFHTLS). The new optical photometry
provided by the Dark Energy Survey (DES) \citep{Abbott2005} provides one
notable advantage over most of these previous optical surveys from the point of
view of high redshift quasar selection. The Dark Energy Camera (DECam) CCD
detectors are optimised to extend to the near infrared wavelengths of
$\sim$1$\mu$m.  The associated $grizY$ filter-set therefore allows detection of
quasars out to higher redshifts than was possible with previous optical
surveys. The DES $z-$band and the redder $Y-$band coupled with the large area
of the celestial sky not previously surveyed and the increased depth over
previous surveys enables detection of Ly$\alpha$ at $6.5 \lesssim z  \lesssim
7.2$. In previous optical surveys the sensitivity in the z band resulted in
  making it increasingly difficult to find quasars with increasing redshift. 

Here we present the first search for high-redshift quasars in the DES Science
Verification Data. DES magnitudes, near infrared (NIR) VISTA magnitudes and
WISE magnitudes are quoted on the AB system.  The conversions from
Vega to AB that have been used for the VISTA data are: $J_{AB} = J_{Vega}
+ 0.937$ and $Ks_{AB} = Ks_{Vega} + 1.839$, these are taken from the Cambridge
Astronomical Survey Unit's website.
\footnote{http://casu.ast.cam.ac.uk/surveys-projects/vista/technical/filter-set}
The conversions for the ALLWISE data are $W1_{AB} = W1_{Vega} + 2.699$ and
$W2_{AB} = W2_{Vega} + 3.339$ which are given in \citet{Jarrett2011} and in the
\textit{ALLWISE} explanatory supplement. \footnote{The \textit{ALLWISE}
explanatory supplement,
http://wise2.ipac.caltech.edu/docs/release/allwise/expsup/sec5\_3e.html,
directs the reader to the \textit{WISE All-Sky} explanatory supplement for the
conversions;
http://wise2.ipac.caltech.edu/docs/release/allsky/expsup/sec4\_4h.html\#summary.}
When required a flat cosmology with $\Omega_{\mathrm{m0}} = $ 0.3 and
H$_{\mathrm{0}} = $ 70.0 was used.

\section{Dark Energy Survey Data}

The Dark Energy Survey is a 5000 deg$^2$ optical survey in the southern
celestial hemisphere being conducted using the Dark Energy Camera (DECam)
\citep{Flaugher2012}. DECam is mounted on the Blanco 4-meter telescope at the
Cerro Tololo Inter-American Observatory (CTIO) and the data are processed by
the DES data management system (\citealt{Mohr2012, Desai2012}).  The 570
Megapixel DECam has a field of view of 3 deg$^2$ with 0.27" per pixel. To allow
for a good detection efficiency of high redshift objects, DECam CCDs are very
sensitive to the red part of the spectrum. These fully depleted, 250
$\mathrm\mu$m thick detectors have been developed at the Lawrence Berkeley
National Laboratory to be used for DES. The quantum efficiency of these
devices in the z band is above 50\%, almost an order of magnitude higher
than traditional thinned devices.  This also allows Y band observations to
be taken within the same survey reducing the need for follow-up photometry to
confirm high redshift quasar candidates.  The combination of depth, red
sensitivity and area provided by DES, makes it an ideal experiment for
searching for luminous quasars at high redshift in the southern celestial
hemisphere.

The DES survey area also overlaps the near infrared VISTA Hemisphere Survey
(VHS) \citep{McMahon2013} at $\sim$1-2$\mu$m and the all-sky Wide Infrared Survey Explorer
(\textit{WISE}) data at 3.4, 4.6, 12 and 22$\mathrm\mu$m. These two surveys
therefore provide infrared photometry for high-redshift quasar candidates
selected in DES \citep{Banerji2015}.

\begin{figure} \includegraphics[width = \linewidth]{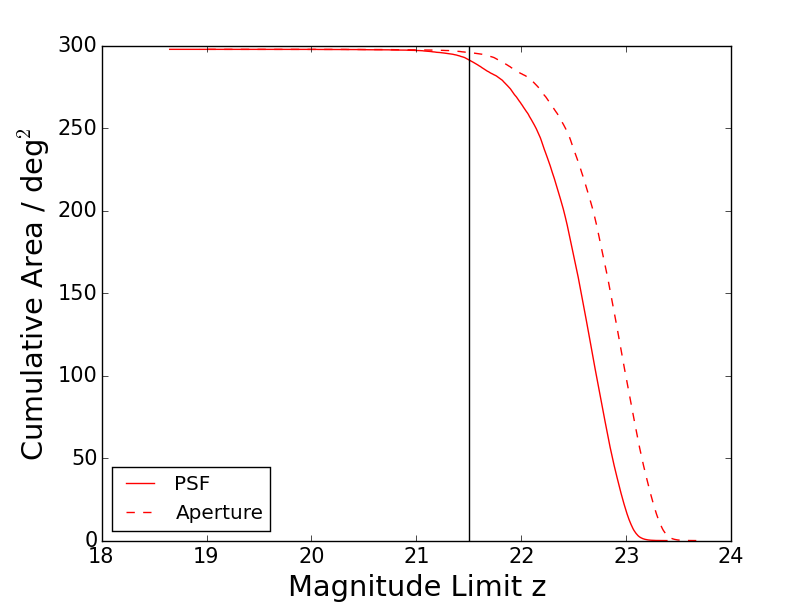}
\caption{Cumulative area versus 10$\sigma$ $z$-band depth 
in a 2$\arcsec$ diameter
aperture and PSF magnitudes for the DES SVA1 data. The
aperture magnitudes were converted to PSF magnitudes using the median offset
between PSF magnitude and aperture magnitude for point sources from the
whole SVA1 dataset. Our magnitude limit of $z_{psf} < 21.5$ is shown as 
the vertical line.
The aperture magnitude limits were taken from the DES Mangle
\citep{Swanson2008} products.} 
\label{zMagLim}
\end{figure}

The DECam obtained its first light images in September 2012 and this was
followed by a period of Commissioning and Science Verification observations (SV
hereafter) lasting between October 2012 and February 2013. In this work we
utilise the Science Verification First Annual Data Release (SVA1) by DES Data
Management to conduct our high-redshift quasar search. The SVA1 observations were
designed to reach the full DES survey depth expected after the nominal 5 year
survey period. However, the depth across the SVA1 area is non-uniform. Figure
\ref{zMagLim} shows the cumulative area in the DES SVA1 region down to a given
10$\sigma$ depth in the $z$-band derived from mangle \citep{Swanson2008} 
based polygon depth maps.
These 10$\sigma$ depth maps are calculated for a 2" diameter apertures. These values will
depend on the seeing in each polygon and have been converted to PSF magnitudes
using the median difference for all stellar objects (as defined in section X) 
The median of these differences ($z_{\mathrm{psf}} - z_{\mathrm{2"}}$) was
found to be -0.28 with a $\sigma_{\mathrm{MAD}} = 0.066$. This offset was
applied to the magnitude limit calculated as shown in figure \ref{zMagLim}. In
this paper PSF and model magnitudes are used. The PSF magnitude is the point
spread function magnitude and the model magnitude is the magnitude once an
galaxy model has been fitted to pick up the flux in the extended wings of the
object.  Both of these parameters are produced by the DES data management
pipeline. As detailed later, a flux-limited sample down to a depth of
$z_{\mathrm{psf}} < 21.5$ is used for the high-redshift quasar search presented
here, which corresponds to a total survey area of 291 deg$^2$. 

\section{Quasar Candidate Selection} \label{SelectionCriteria}

Whilst low redshift ($z\la2$) quasars have characteristically blue optical
colours (from ultra violet excess) \citep{Schmidt1983} at $z\ga4$ quasars start
to become red in the shorter wavelength optical bands (\citealt{Hook1998,
StorrieLombardi2001}).  The observer frame optical colours of these high
redshift quasars are characterised by absorption due to neutral hydrogen in the
intergalactic medium in the rest frame UV shortward of the Lyman-$\alpha$
($\lambda_{rest}=1216\rm\AA$) emission line. At z $=$ 6 the Ly$\alpha$ line is
redshifted to 8512$\rm\AA$ which leads to red $(i-z)$ colours for quasars at
these redshifts.  These high-redshift quasars are therefore selected in optical
surveys as $i-$band drop outs \citep{Fan2000}. Below we detail our selection
method for isolating high-redshift quasar candidates from the DES imaging data.
PSF magnitudes are used throughout (unless otherwise stated) as they provide
the best measure of flux for unresolved point-like sources.

\begin{figure} \includegraphics[width = \linewidth]{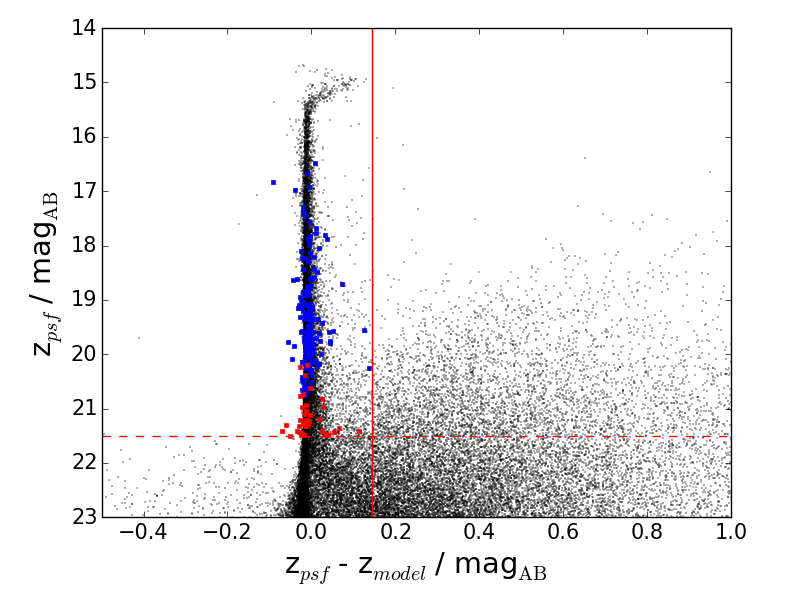}
\caption{The black points are all the objects on one tile from the survey and
the red points show the quasar candidates after the final stage of the
selection in section \ref{SelectionCriteria}. A clear locus can be seen at 0.0
on the horizontal axis. The blue points are known quasars taken from the
catalogue given in \citet{Veron2010} with $z > 2.0$. Below this redshift host
galaxy contamination can occur. The median of the Veron 2010 sample is 0.015
and the $\sigma_{\mathrm{MAD}}=$ 0.052. The solid vertical line shows the cut
used to select point sources. The dashed horizontal line shows the magnitude
limit used.}
\label{SG} \end{figure}

The primary colour selection used is an $(i-z)$ colour criterion, which is
based on the selection used by \citet{Fan2001}. The \citet{Fan2001} SDSS
selection of $(i-z)>2.2$ was mapped onto the DES filter system as follows.
First we selected point-like objects in the area of overlap between the DES and
SDSS surveys, as those having $z_{\mathrm{psf}} - z_{\mathrm{model}} < 0.145$. This
point source selection is illustrated in figure \ref{SG} and is detailed below.
The DES and SDSS sources were matched using a radius of 2.16$\arcsec$
(0.006$\degree$).  
A least squares linear fit was carried out on a running median of the colours.
A window size of 50 was used for the median and the data was sorted by $(i
- z)_{\mathrm{SDSS}}$ and then sparsely sampled along the $(i
- z)_{\mathrm{SDSS}}$ direction with a stride of 10 to determine the
colour-term between the two surveys:  \begin{equation} (i- z)_{\mathrm{DES}}
= 0.72(i - z)_{\mathrm{SDSS}} + 0.111 \end{equation} Based on this and the
\citet{Fan2001} selection criterion we derive an equivalent colour criterion of
$(i - z)_{\mathrm{DES}} > 1.694$.

We now list each of the steps in our high-redshift quasar selection process,
which are also summarised in Table \ref{tab:selection}.

\begin{figure} \includegraphics[width = \linewidth]{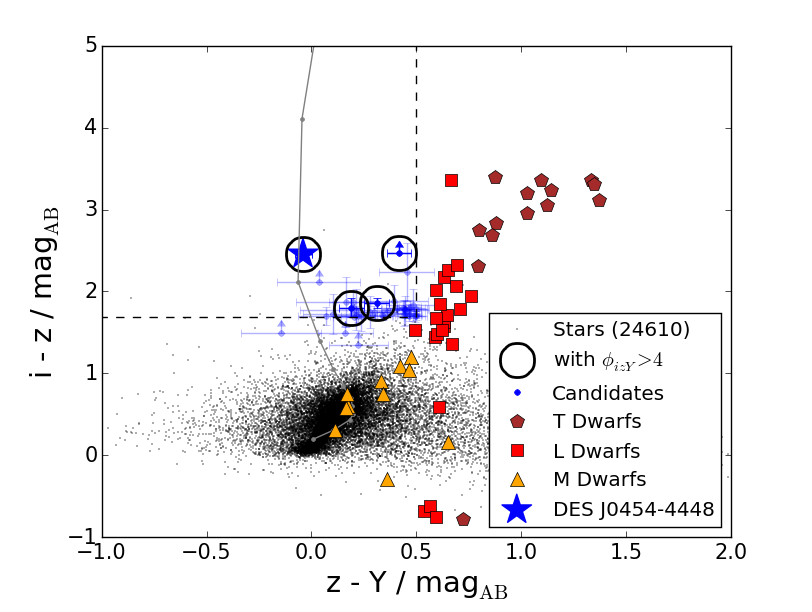}
\caption{The black points represent all objects that have $z_{\mathrm{psf}}
- z_{\mathrm{model}} < 0.145$ and are detected in the i, z and Y bands from one
tile of the survey.  The blue points are the candidate objects that satisfy the
first 11 steps of the selection criteria. The objects that pass the final step
are circled while the others are faded. The blue star shows the location of the
confirmed quasar.  The grey line shows a simulated quasar track starting from
$z = 5$ and incrementing by 0.1 to $z = 6.4$. Objects that have i = 99 (are
undetected in the i band) pass the selection criteria and are here plotted with
arrows to indicate that they are lower limits in i-z. The objects outside the
selection box are limits in the i band where the survey is particularly
shallow. Many of the objects are situated at the lower right of the selection
box which suggests that they are cool stars scattered in by random photometric
errors. This is supported by the large errors that many of them have which
overlap the selection box edges. The red, orange and brown points are known
brown dwarfs found with UKIDSS and converted using the spectra and the DES band
passes.} 
\label{CandidatesGraph}
\end{figure}

\begin{enumerate} \item  High-redshift quasar and other rare object searches
are often contaminated by a large number of artefacts. These spurious sources
often dominate outlier regions of colour space that are also populated by
smaller numbers of real objects. In fact it is safe to assume that most
outliers in an unfiltered catalogue are junk.  The first step therefore involved
artefact removal at the catalogue level.  Objects that were flagged by the DES
data management SVA1 pipeline as having bright neighbours that were close enough
to bias photometry, being originally blended with another object or having at
least one saturated pixel were provisionally labelled as junk. If they were also
flagged as having at least one low-weight pixel in the measurement image within
the isophotal footprint or in the filtered detection image then this was
confirmed and they were removed.  A combination of the weight flags and the
image flags was used to ensure that faint objects undetected in some of the
bands were retained while removing bright saturated objects and bleed trails.
The flags cut resulted in 36,959,467 objects remaining of the original
40,129,963 in the DES SVA1 release. Examples of the artefacts removed in this
step are shown in Appendix \ref{JunkRemoval}.


\item A flux limited sample was then created by limiting the sample to sources
with: \begin{equation} z_{\rm{psf}} \leq 21.5\ {\rm and}\  \sigma_{z} < 0.1
\end{equation}

\item Objects were required to satisfy a point source criteria of:
\begin{equation} z_{\mathrm{psf}}- z_{\mathrm{model}} < 0.145 \end{equation}
which is the star-galaxy classifier defined in the SDSS photometric survey
(\citet{Stoughton2002}) and SDSS quasar selection (\citet{Richards2002}). After
this step there were 4,790,106 objects remaining.  This star-galaxy separation
criterion is shown in figure \ref{SG}.  The stellar locus is clearly situated
at $z_{\mathrm{psf}} - z_{\mathrm{model}} \sim 0$. We can see that the
star-galaxy separation starts to become less reliable at fainter magnitudes but
the stellar locus is still clearly visible down to $z<21.5$, which is the flux
limit imposed on our sample. 

\item The colour criterion derived above: \begin{equation} (i-z)_{\mathrm{DES}}
> 1.694 \end{equation} was then applied leaving 23,517 objects.

\item As high redshift ($z\ga6$) quasars should not have detectable flux in
their rest frame shortward of the hydrogen limit limit at
$\lambda_{\mathrm{rest}}= 912\AA$ corresponding to $\lambda_{\mathrm{obs}}\sim
6500\AA$ they should be undetected in the $g$ or $r$ bands.  Visual inspection
of a sample of candidates indicated that some had significant detectable flux
in $g$ and $r$. We therefore applied the following criterion to the $g$ and $r$
band measurements \begin{equation} g_{\rm{psf}} > 23.0\ {\rm and}
\ r_{\rm{psf}} > 23.0 \end{equation} which were chosen as this ruled out almost
all of the objects that were reliably detected without removing undetected
sources in regions of particularly deep photometry.

\item There were a few objects that were very faint in the g or r band but were
in a very deep part of the survey and still had a reliable detection in these
bluer bands. To remove these detected objects an error threshold of
$\sigma_{g}$ and $\sigma_{r} >$ 0.1 was used. i.e. Objects with $\sigma_{g}$ and
$\sigma_{r} \sim 0.2$ (5$\sigma$ detections) are retained since there is a 
finite probability that they have zero flux.

\item Cool dwarf stars, such as L and T-dwarfs, in our own Milky Way, are the
main astrophysical contaminants in high redshift quasar searches. These were
removed using a colour cut of \begin{equation} z_{\mathrm{psf}}
- Y_{\mathrm{psf}} < 0.5 \end{equation}

The colour selection can be seen in Figure \ref{CandidatesGraph} which shows
the cool dwarf locus clearly distinct from the selection box in terms of the
$z-Y$ colour. We note that a blue $z-Y$ cut will also potentially remove higher
redshift quasars at $z \gtrsim 6.5$, at which point the Ly$\alpha$ peak begins
to move into the DES $Y$-band, making the $z-Y$ colour redder. In the future,
we will use NIR data, particularly in the $J$-band from VHS, to push our quasar
search to higher redshifts. This selection will cover $z-Y$ dropouts, rather
than $i-z$, and will be discussed further in future work.

\item At this stage many of the objects left were bright artefacts in the
z band which were not present in any of the other bands. To limit the numbers
of these that was present in the sample two Y band magnitude cuts were used.
First objects with bad Y band photometry (Y$_{\mathrm{psf}} = 99$) were removed
and then those with Y$_{\mathrm{psf}} < 23.0$. These cuts ensured that objects
were detected in the Y band as well reducing their likelihood of being a z band
artefact.

\item Following these cuts, many of the objects that remained had PSF
magnitudes of 99 in the $r$-band.  They corresponded to two classes of objects:
either very bright objects or those at the $r$-band detection limit. The bright
objects were removed using list driven forced photometry in the $r$-band. For
each object we identified the brightest pixel in a box of 3x3 pixels centred on
the position of the object and required that it was no more than 10 analog to
digital units greater than the median value in a 30$\arcsec$ box around the
object. This corresponds roughly to three times the median
$\sigma_{\mathrm{MAD}}$ (3.06) above the median of the median values (0.147) of
the images in the data release. A conservative threshold of 10 was used to
filter these bright objects to ensure that any faint objects which may
correspond to high-redshift quasars, were not removed.

\item Most of the objects that remained at this stage were cosmic ray
detections in the $z$-band.  These were removed based on the differences
between adjacent pixels (as a proxy for the local pixel gradients) as a cosmic
ray would be more abrupt in terms of its pixel gradient than a real object. In
future DES data releases, a new method of cosmic ray reduction will be included
in the pipeline processing.

\item Following these steps there were few enough candidates to cut down by
systematic visual inspection. The visual inspection stage removed 34 spurious
sources such as cosmic rays, satellite trails and diffraction spikes. At the
end of this stage we have 43 candidate high-redshift quasars. 

\item We ranked these 43 final candidates based on an error weighted
photometric selection metric for follow-up spectroscopy. The selection metrics
$\phi_{\mathrm{iz}}$ and $\phi_{\mathrm{zY}}$ combine to give
$\phi_{\mathrm{izY}}$ which essentially constitutes a error-weighted measure of
how far into the colour-selection box, shown in Figure \ref{CandidatesGraph},
each candidate lies. 

\begin{align} 
\label{Phi_eq}
\phi_{\mathrm{iz}} &= \frac{(i_{\mathrm{psf}} - z_{\mathrm{psf}})
- 1.694)}{\sqrt{i_{\mathrm{err}}^{2} + z_{\mathrm{err}}^{2}}} \\ 
\phi_{\mathrm{zY}} &= \frac{0.5 - (z_{\mathrm{psf}}
- Y_{\mathrm{psf}})}{\sqrt{z_{\mathrm{err}}^{2} + Y_{\mathrm{err}}^{2}}} \\
\phi_{\mathrm{izY}} &=
\sqrt{\frac{\phi_{\mathrm{iz}}}{\vert\phi_{\mathrm{iz}}\vert}\phi_{\mathrm{iz}}^{2}
+ \frac{\phi_{\mathrm{zY}}}{\vert\phi_{\mathrm{zY}}\vert}\phi_{\mathrm{zY}}^{2}} 
\end{align}

\noindent 

For those sources undetected in the $i$-band, the $i$-band magnitude limit at
the source position was used instead in the computation of $\phi_{\mathrm{iz}}$ and the
error set to 0.2. 
\end{enumerate}

\begin{figure} \includegraphics[width = \linewidth]{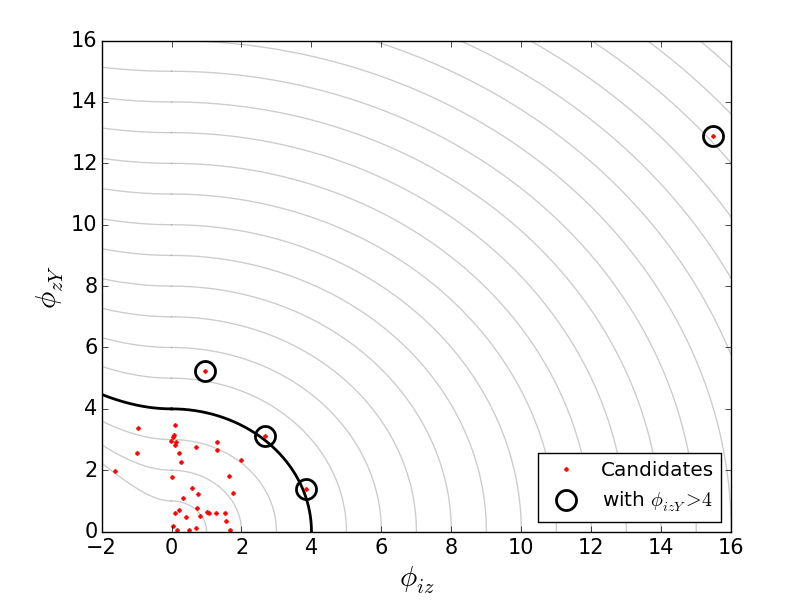} 
\caption{The $\phi_{\mathrm{izY}}$ selection metric. Here the x axis is the
$\phi_{\mathrm{iz}}$ defined in equation \ref{Phi_eq} and the y axis is $\phi_{\mathrm{zY}}$. The four
objects with $\phi_{\mathrm{izY}} >$ 4 are circled and the object that was confirmed as
a quasar has the highest value of both $\Phi_{iz}$ and $\Phi_{zY}$ the figure.} 
\label{Phi_izy} 
\end{figure}

\begin{figure*} \includegraphics[width = \linewidth]{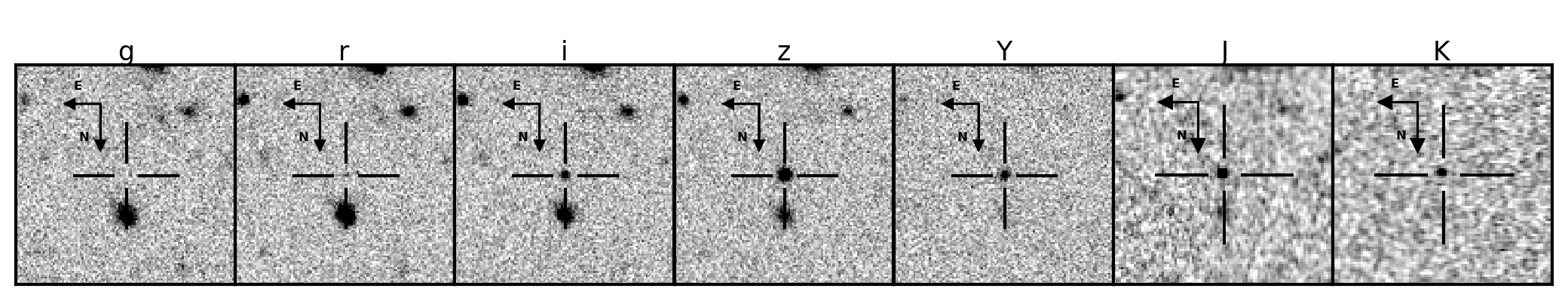}
\caption{Cutouts of our spectroscopically
confirmed quasar, DESJ0454-4448 in each of the DES (g, r, i, z and Y) and VHS (J
and K) wavebands. The boxes are 30" across and the quasar is the central
object.} \label{Cutouts} \end{figure*}

This selection led to prioritising four of the 43 candidates as highly viable
high-redshift quasar candidates with $\phi_{\mathrm{izY}} > 4$ from the
291 deg$^2$ region. The distribution of $\phi$ values can be seen in Figure
\ref{Phi_izy} where the four candidates have been circled. One of these
candidates clearly has much larger values of $\phi$ than the other three and
this object (DES J0454$-$4448) with $z_{\mathrm{AB}}$ = 20.2 was therefore
selected for follow-up spectroscopy as our most promising first high-redshift
quasar candidate from the DES data. DES J0454$-$4448 is shown across the DES
and VISTA bands in figure \ref{Cutouts}. In these cutout images the absence of
flux in the g and r bands can be clearly seen along with the increase in
brightness from the i to z bands.

\begin{table} \begin{center} \caption{Summary of the steps in the high-redshift
quasar selection process} \label{tab:selection} \begin{tabular}{cccc} \hline
Step & Description & Number& Number\\ &  & Removed & Remaining \\ \hline
& Number of objects in database & & 40,129,963  \\
1 & Flag criteria & 3,170,496 & 36,959,467 \\
2 & z$_{\mathrm{psf}}\leq$ 21.5 and $\sigma_{\mathrm{z}} < 0.1$ & 29,760,121
  & 7,199,346 \\
3 & $z_{\mathrm{psf}} - z_{\mathrm{model}} < 0.145$ & 2,409,240 & 4,790,106 \\
4 & i$_{\mathrm{psf}}$ - z$_{\mathrm{psf}} >$ 1.694 & 4,766,589 & 23,517 \\
5 & g$_{\mathrm{psf}}$ and r$_{\mathrm{psf}} >$ 23.0 & 12,985 & 10,532 \\
6 & $\sigma_{g}$ and $\sigma_{r} >$ 0.1 & 371 & 10,161 \\
7 & z$_{\mathrm{psf}}$ - Y$_{\mathrm{psf}} <$ 0.5 & 92 & 10,069 \\
8a & Y$_{\mathrm{psf}}$ not equal to 99 & 8725 & 1344 \\ 
8b &  Y$_{\mathrm{psf}}  < 23.0$ & 506 & 838 \\
9 & Forced photometry in r band & 372 & 466 \\
10 & Cosmic ray removal & 489 &  77 \\
11 & Visual Inspection & 34 & 43 \\
12 & $\phi_{\mathrm{izY}} >$ 4 & 39 & 4 \\ \hline \end{tabular} \end{center} \end{table}

\section{DES J0454$-$4448}

\begin{figure} \includegraphics[width = \linewidth]{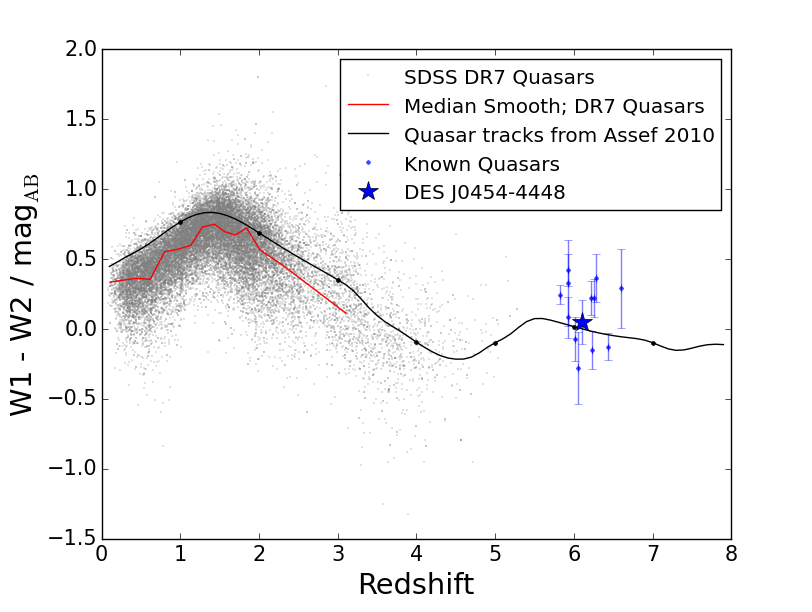} 
\caption{The figure shows the evolution in W1-W2 colour with redshift of known
quasars. The grey points are known low-redshift quasars from SDSS DR7 with the
red line being a median smooth of these data. This allows comparison with the
predicted track, from \citet{Assef2010}, which is shown by the black line. The
blue points are a sub sample of the known high redshift quasars matched to
WISE, these objects were matched to the ALLWISE data and then visually
inspected to remove blended sources in WISE. The large blue star is DES J0545-4448.} 
\label{WISE_zred_W1W2} 
\end{figure}

The photometric data for DES J0454$-$4448 is summarised in Table
\ref{tab:properties}. This includes the DES photometry (g, r, i, z and Y) as
well as infrared magnitudes from the VISTA Hemisphere Survey (J and K) and the
\textit{AllWISE} data release of the \textit{WISE} All-Sky Survey
\citep{Wright2010} (W1 and W2). \citet{Blain2013} have studied the infrared
\textit{WISE} colours of known high-redshift quasars finding that they are
typically bluer than the W1$_{\mathrm{AB}}$ - W2$_{\mathrm{AB}}$ $>$ 0.16
selection cut used to select lower redshift AGN such as in \citet{Stern2012}.
The colour (W1$_{AB}$ - W2$_{AB}$ $=$ 0.05) of this object supports this and
agrees with the locus of known high redshift quasars matched to the WISE data.
The colour also agrees with the predicted track in \citet{Assef2010} as shown
in Figure \ref{WISE_zred_W1W2}.  As the resolution of WISE (6.1" and 6.4" in W1
and W2 respectively) is quite large compared to the separation of the object
and its nearest neighbour (4.89") it is debatable if the magnitude corresponds
solely to the quasar.

\begin{figure*} 
\includegraphics[width = \linewidth]{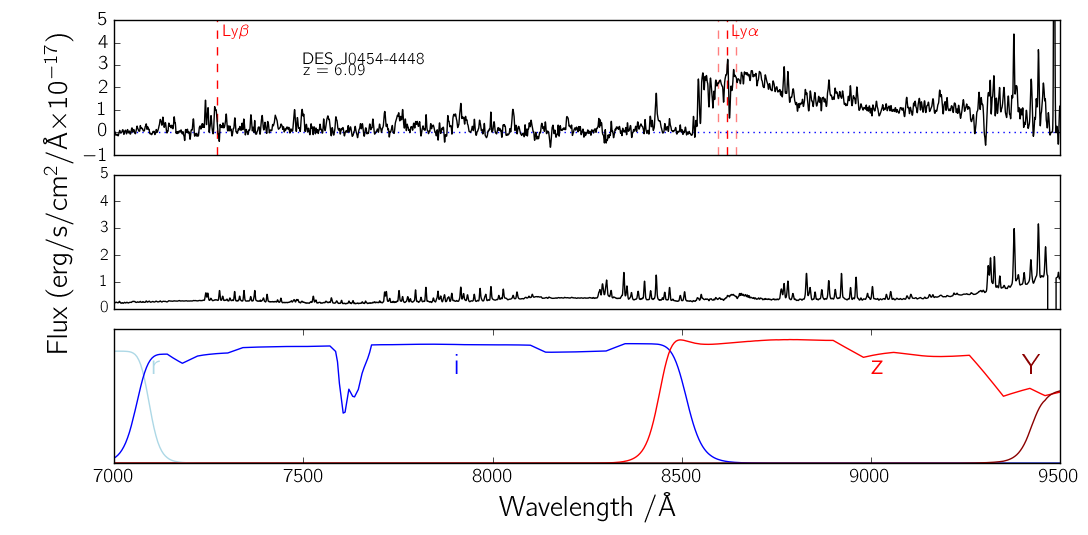}
\caption{The discovery spectrum of DES J0454-4448 at z = 6.09. It shows the
characteristic flux deficit blueward of Lyman alpha indicative of a high
redshift quasar. The red vertical lines over the spectrum show the positions of
the Ly-$\beta$ and the Ly-$\alpha$ emission line redshifted to z $=$ 6.09 as
well as the errors on the position of Ly-$\alpha$ from the error in the
redshift. The middle plot shows the error spectra and the bottom plot indicates
the efficiencies of the DES filters as a function of wavelength. Both the
object spectra and associated error spectra have been smoothed to the same
degree using a gaussian filter.}
\label{Spectra} \end{figure*}

Spectroscopic follow-up of DESJ0454$-$4448 was conducted using the Magellan
Echellete (MagE) spectrograph \citep{Marshall2008} on the 6.5m Clay
telescope at the Las Campanas Observatory in Chile.  Two 10 minute exposures
were taken on the night of December 2nd 2013 in variable conditions.  The
data were reduced and the one-dimensional spectrum combined  using a custom
set of {\sc idl} routines.  The discovery spectrum for the quasar can be
seen in Fig. \ref{Spectra}.

Measuring the quasar redshift from emission lines in the present spectrum is
difficult due to the strong attenuation of Ly$\alpha$ 
($\lambda_{rest}$=1215.67\AA) and the modest quality and limited
wavelength range of the data which does not cover the
wavelenght range to detect CIV(1549) or SiIV+OIV](1400)
which at z $\geq$ 6.00 have observed wavelenghthat $\lambda$ $\geq$ 9800\AA.
We therefore rely on the apparent onset of Ly$\alpha$ forest
absorption to estimate the redshift. This approach has been previously used for
weak lined quasars such as BL Lac objects \citep{Danforth2010}.

A series of absorption lines appears
over $8530 \lesssim \lambda \lesssim 8630$~\AA, which we interpret as the
expected Ly$\alpha$ absorption in the quasar's near zone.  The reddest
of these lines  falls at 8626~\AA, which corresponds to a Ly$\alpha$
redshift of $z = 6.096$.  We tentatively detect the quasar’s Si II
1260~\AA\ emission line with a peak wavelength near 8920~\AA, though it is
contaminated by skyline subraction residuals, corresponding to a redshift
of 6.077. Combining these estimates gives a mean redshift and 
standard deviation: $z = 6.09 \pm 0.01$. To reflect the systematic
uncertainty in the redshift we follow \citet{Fan2006a}
and adopt an total uncertainty of 0.02 to take account of the observed
offsets in redshifts between low and high ionization lines 
in quasars.

\begin{figure} 
\includegraphics[width = \linewidth]{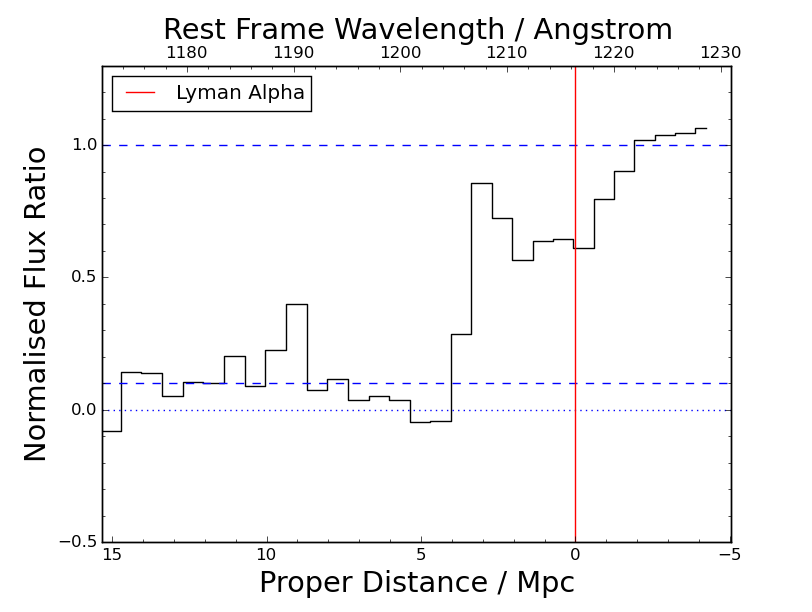} 
\caption{In this plot the ratio of the continuum flux and the spectrum is shown.
The red vertical line shows the position of the Lyman alpha peak. The dashed
blue line is at a flux ratio of 0.1 and the dotted blue line show the zero
level. The nearzone has a size of $4.1_{-1.2}^{+1.1}$ proper Mpc.} 
\label{NZPlot} 
\end{figure}

The quasar continuum is strongly attenuated at $\lambda \le 8530$~\AA.
Following the method outlined in \citet{Fan2006b}, we take the edge of the
proximity zone to be the point where the ratio between the continuum flux
and the spectrum first falls below 0.1 blueward of the Ly$\alpha$ peak. 
Here the intrinsic continuum is roughly estimated using a composite from
low-redshift SDSS spectrum (as used in \citet{Mortlock2011}) normalized to the
flux between 1268 and 1295\AA.  The nearzone size for an edge at
8530~\AA\ and the quasar redshift estimated above is $4.1_{-1.2}^{+1.1}$ proper
Mpc, the error in this measurement is dominated by the error in the
redshift.  In figure \ref{NZPlot} the normalised flux and the extent of the
nearzone are shown. The corrected nearzone size of this object is
found to be consistent with the measured relationship in the \citet{Carilli2010}
study. Due to the low signal to noise of the data no constraints can be put on
the optical depth at this redshift.

\begin{table} \begin{center} \caption{Properties of DES J0454-4448. The g band
magnitude is given as a 10$\sigma$ magnitude limit for a 2" aperture.}
\label{tab:properties} \begin{tabular}{cc}
\hline 
& DES J0454-4448 \\ 
\hline 
DES Tilename & DES0453-4457 \\
RA (J2000) &  73.50745 (04$^{h}$54$^{m}$01.79$^{s}$)\\ 
DEC (J2000) &  -44.80864 (-44$^{\circ}$48'31.1")\\ 
Redshift & 6.09 $\pm$ 0.02\\
g & $>$ 23.41\\
r & 25.30 $\pm$ 0.25 [AB]\\ 
i & 22.66 $\pm$ 0.05 [AB]\\
z & 20.20 $\pm$ 0.01 [AB]\\ 
Y & 20.24 $\pm$ 0.04 [AB]\\ 
J & 20.24 $\pm$ 0.07 [AB]\\ 
Ks & 20.12 $\pm$ 0.17 [AB]\\ 
W1 & 19.68 $\pm$ 0.08 [AB]\\ 
W2 & 19.62 $\pm$ 0.14 [AB]\\ 
M$_{\mathrm{1450}}$ & -26.48 $\pm$ 0.03 \\
R$_{NZ}$ & 4.1 $_{-1.2}^{+1.1}$ Mpc \\
R$_{NZ, corrected}$ & 4.8$_{-1.4}^{+1.3}$ Mpc \\
\hline 
\end{tabular} 
\end{center} 
\end{table}

\section{Discussion}

The number density estimates based on the \citet{Willott2010} luminosity function
suggest that there will be between 50 and 100 new quasars with z $>$ 6.0 across
the $\sim$5000 deg$^{2}$ of the full DES area down to Y$_{AB}$ = 21.5. The
predictions also suggest that there will be 3 to 10 quasars with z $>$ 6.5 and
Y$_{AB}<$ 21.5 found in the complete survey footprint.

Estimates from the \citet{Willott2010} and \citet{McGreer2013} luminosity
functions suggest that there are 5 to 10 quasars with z $<$ 6.5 in the SVA1
area. Our selection method gives 43 other candidate objects and suggests four
as higher priority than the others. This final selection step is biased towards
bright objects, due to them having smaller errors than fainter candidates, and
is not designed to find a complete sample.  To find all the quasars in this
area with a high purity would require full Bayesian analysis of the quasar
population and the brown dwarfs. The exact number of quasars predicted in our
survey is also sensitive to the choice of quasar template. Different
predictions for the Ly$-\alpha$ forest in these quasar template spectra can
change the exact redshift range probed by our colour selection criteria. This
then heavily influences the predicted number of objects.

\section{Conclusions}

We have presented the first search for high-redshift quasars in the Dark Energy
Survey using 291 deg$^2$ of optical imaging data from the DES SV observations.
Our search algorithms are demonstrated to be effective in removing image
artefacts and spurious sources such as cosmic rays and satellite trails that
can populate the same regions of colour space as rare objects in the data.
Based on the DES photometric data alone, we were able to reduce the number of
candidates in this region to a size suitable for visual inspection. We also did
not require confirmation photometric follow-up as has been the case in previous
high-redshift (z $>$ 6) quasar searches.  A photometric error-weighted colour
selection metric was introduced as a method of ranking candidates according to
the probability of being high-redshift quasars. The highest ranked
candidate, DESJ0454$-$4448 was spectroscopically confirmed to be a high
redshift quasar at $z=6.10$. Three more candidates were also identified as
having colours consistent with high-redshift quasars and will be further
investigated as part of the larger sample of high-redshift quasar candidates
from the DES Year 1 observations. 

DESJ0454$-$4448 is also detected at infrared wavelengths in the VISTA
Hemisphere Survey and the \textit{WISE} All-Sky Survey and its infrared colours
are entirely consistent with those expected for high redshift quasars. Using
current estimates of the $z\sim6$ quasar luminosity function, we have shown
that the first $z\sim6$ quasar is expected in the 291 deg$^2$ region used in
this work, at a $z$-band magnitude of 20.2. The measured $z$-band magnitude of
DESJ0454$-$4448 is 20.2, entirely consistent with these predictions. 

We conclude that our search algorithms for high-redshift quasars are successful
in isolating quasars at $z>6$ from DES photometry alone. The analysis method
presented here should be easily scalable to the full 5000 deg$^2$ area of DES
and is therefore expected to result in substantial numbers of new high-redshift
quasars in the southern hemisphere over the next few years.

\section{Acknowledgements}
RGM, SLR, MB acknowledge the support of UKScience and Technology research
Council (STFC).

We are grateful for the extraordinary contributions of our CTIO colleagues and
the DES Camera, Commissioning and Science Verification teams in achieving the
excellent instrument and telescope conditions that have made this work
possible.  The success of this project also relies critically on the expertise
and dedication of the DES Data Management organization.

Funding for the DES Projects has been provided by the US Department of Energy,
the US National Science Foundation, the Ministry of Science and Education of
Spain, the Science and Technology Facilities Council of UK, the Higher
Education Funding Council for England, the National Center for Supercomputing
Applications at the University of Illinois at Urbana-Champaign, the Kavli
Institute of Cosmological Physics at the University of Chicago, Financiadora
de Estudos e Projetos, Funda{\c c}{\~a}o Carlos Chagas Filho de Amparo {\'a}
Pesquisa do Estado do Rio de Janeiro, Conselho Nacional de Desenvolvimento
Cient{\'i}fico e Tecnologico and the Minist{\'e}rio da Ci{\^e}ncia
e Tecnologia, the Deutsche Forschungsgemeinschaft and ˆ the Collaborating
Institutions in the Dark Energy Survey.

The Collaborating Institutions are Argonne National Laboratories, the
University of California at Santa Cruz, the University of Cambridge, Centro de
Investigaciones Energeticas, Medioambientales y Tecnologicas-Madrid, the
University of Chicago, University
College London, the DES-Brazil Consortium, the Eidgenossische Technische
Hochschule (ETH) Zurich, Fermi National Accelerator Laboratory, the University
of Edinburgh, the University of Illinois at Urbana-Champaign, the Institut de
Ciencies de l’Espai (IEEC/CSIC), the Institut de Fisica d’Altes Energies, the
Lawrence Berkeley National Laboratory, the Ludwig-Maximilians Universit{\"a}t
and the associated Excellence Cluster Universe, the University of Michigan, the
National Optical Astronomy Observatory, the University of Nottingham, The Ohio
State University, the University of Pennsylvania, the University of Portsmouth,
SLAC National Laboratory, Stanford University, the University of Sussex, and
Texas A\&M University.

This paper has gone through internal review by the DES collaboration.

The analysis presented here is based on observations obtained as part of the
VISTA Hemisphere Survey, ESO Progamme, 179.A-2010 (PI: McMahon).

ACR acknowledges financial support provided by the PAPDRJ CAPES/FAPERJ
Fellowship

\bibliographystyle{mn2e}
\bibliography{Refs}

\section*{Affiliations}
{\small
$^{1}$Institute of Astronomy, University of Cambridge, Madingley Road, Cambridge CB3 0HA, UK\\
$^{2}$Kavli Institute for Cosmology, University of Cambridge, Madingley Road, Cambridge CB3 0HA, UK\\
$^{3}$Department of Physics \& Astronomy, University College London, Gower Street, London, WC1E 6BT, UK\\
$^{4}$Center for Cosmology and Astro-Particle Physics, The Ohio State University, Columbus, OH 43210, USA\\
$^{5}$Department of Astronomy, The Ohio State University, Columbus, OH 43210, USA\\
$^{6}$Space Telescope Science Institute, 3700 San Martin Dr, Baltimore, MD 21218, USA\\
$^7$Observatories of the Carnegie Institution for Science, 813 Santa Barbara Street, Pasadena, CA 91101, USA\\
$^8$Department of Physics, University of Michigan, Ann Arbor, MI 48109, USA\\
$^9$Cerro Tololo Inter-American Observatory, National Optical Astronomy Observatory, Casilla 603, La Serena, Chile\\
$^{10}$Fermi National Accelerator Laboratory, P. O. Box 500, Batavia, IL 60510, USA\\
$^{11}$Institut d'Astrophysique de Paris, Univ. Pierre et Marie Curie \& CNRS UMR7095, F-75014 Paris, France\\
$^{12}$SLAC National Accelerator Laboratory, Menlo Park, CA 94025, USA\\
$^{13}$Observat\'orio Nacional, Rua Gal. Jos\'e Cristino 77, Rio de Janeiro, RJ - 20921-400, Brazil\\
$^{14}$Institute of Cosmology \& Gravitation, University of Portsmouth, Portsmouth, PO1 3FX, UK\\
$^{15}$George P. and Cynthia Woods Mitchell Institute for Fundamental Physics and Astronomy, and Department of Physics and Astronomy, Texas A\&M University, College Station, TX 77843,  USA\\
$^{16}$Department of Physics, Ludwig-Maximilians-Universit\"at, Scheinerstr. 1, D-81679 M\"unchen, Germany\\
$^{17}$Kavli Institute for Particle Astrophysics \& Cosmology, P. O. Box 2450, Stanford University, Stanford, CA 94305, USA\\
$^{18}$Laborat\'orio Interinstitucional de e-Astronomia - LIneA, Rua Gal. Jos\'e Cristino 77, Rio de Janeiro, RJ - 20921-400, Brazil\\
$^{19}$Institut de Ci\`encies de l'Espai, IEEC-CSIC, Campus UAB, Facultat de Ci\`encies, Torre C5 par-2, 08193 Bellaterra, Barcelona, Spain\\
$^{20}$University Observatory Munich, Scheinerstrasse 1, 81679 Munich, Germany\\
$^{21}$Department of Physics, The Ohio State University, Columbus, OH 43210, USA\\
$^{22}$Australian Astronomical Observatory, North Ryde, NSW 2113, Australia\\
$^{23}$ICRA, Centro Brasileiro de Pesquisas F\'isicas, Rua Dr. Xavier Sigaud 150, CEP 22290-180, Rio de Janeiro, RJ, Brazil\\
$^{24}$Institut de F\'{\i}sica d'Altes Energies, Universitat Aut\`onoma de Barcelona, E-08193 Bellaterra, Barcelona, Spain\\
$^{25}$Brookhaven National Laboratory, Bldg 510, Upton, NY 11973, USA\\
$^{26}$Department of Physics and Astronomy, University of Pennsylvania, Philadelphia, PA 19104, USA\\
$^{27}$Centro de Investigaciones Energ\'eticas, Medioambientales y Tecnol\'ogicas (CIEMAT), Madrid, Spain\\
$^{28}$Instituto de F\'\i sica, UFRGS, Caixa Postal 15051, Porto Alegre, RS - 91501-970, Brazil\\
$^{29}$Department of Physics, Stanford University, 382 Via Pueblo Mall, Stanford, CA 94305, USA\\
$^{30}$SEPnet, South East Physics Network, (www.sepnet.ac.uk)\\
}

\appendix
\section{Exploratory Data Analysis}

Given below are some examples of the types of object removed during each step
of the selection process.

\subsection{Flags}
\label{JunkRemoval}
\begin{figure}
\includegraphics[width = \linewidth]{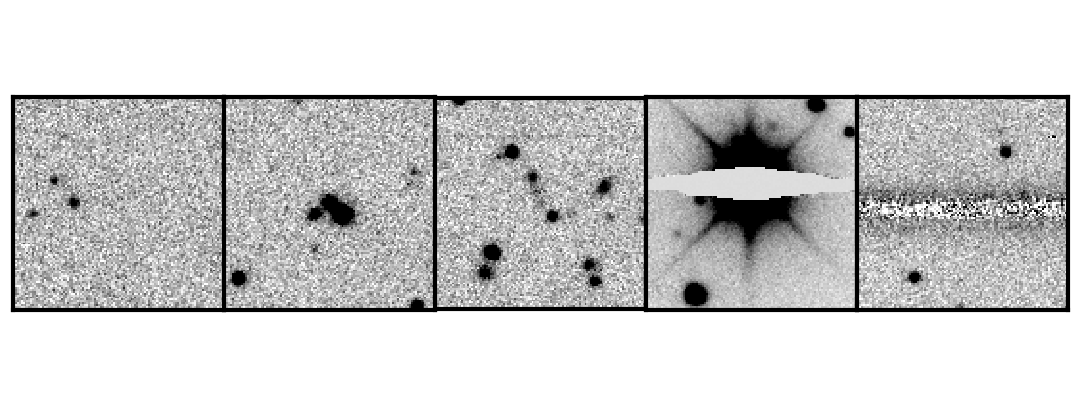}
\caption{All of the objects shown here are flagged with a three in the DES
flags. It can be seen that all of these objects have very different problems
with them and that the first cutout shows an object that we would want to keep
in if it looked like that in the g or r band. These cutouts are 30" by 30" and
all taken from the z band images. The flagging was found to be similar across
all the bands and objects shown here have many analogies in the other bands.}
\label{CutoutsFlag3}·
\end{figure}

The DES flags were used as detailed in step one of section
\ref{SelectionCriteria}. In figure \ref{CutoutsFlag3} we show examples of
objects that fail only the first part of our flag cuts. As can be seen most of
these objects are junk rather than valid candidates. The central cutout shows
an object that is undetected in that band which are exactly the objects that we
want to keep in the g and r bands as they correspond to high redshift quasar
candidates. To ensure that candidates were not removed objects were only
removed if they failed both of the flag cuts. The second cut was made in the
weight flags and the combination of these removed the junk objects but kept the
faint, undetected sources in.

The internal flags that are produced by SExtractor are:\\
1 The object has neighbours, bright and close enough to significantly bias the MAG AUTO
photometry, or bad pixels (more than 10\% of the integrated area affected)\\
2 The object was originally blended with another one\\
4 At least one pixel of the object is saturated (or very close to)\\
8 The object is truncated (too close to an image boundary)\\
16 Object’s aperture data are incomplete or corrupted\\
32 Object’s isophotal data are incomplete or corrupted\\
64 A memory overflow occurred during deblending\\
128 A memory overflow occurred during extraction\\
this information was taken from the SExtractor documentation.
\\
These flags are produced for each band and are included in the DES 
DM database products.

\subsection{g and r Band Detections}

\begin{figure}
\includegraphics[width = \linewidth]{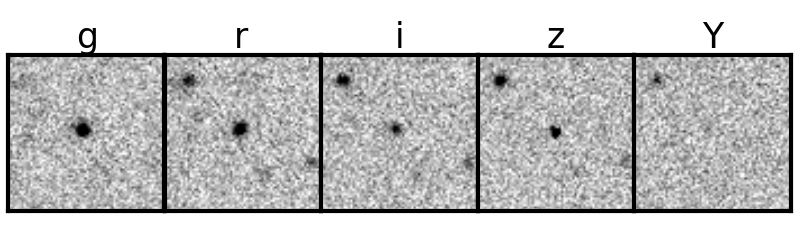}
\caption{In these cutouts an object is shown with $g_{\mathrm{psf}} = 23.16
\pm 0.03$ and $r_{\mathrm{psf}} = 23.02 \pm 0.03$. The source is
obviously detected in the g and r band despite having a magnitude below our
limit.}
\label{Cutoutsgr}
\end{figure}

In steps 5 and 6 a catalogue based method for removing objects that are
detected in the bluer bands is used. First objects with a magnitude brighter
than 23 were removed. It was found that this did not reliably remove all
detected objects. Some objects were in a particularly deep part of the survey
and still had a clean detection, shown in figure \ref{Cutoutsgr}, despite being
fainter than 23rd magnitude. To remove these an error cut was used, sources
with $\sigma_{g/r} > 0.1$ were kept.

\begin{figure}
\includegraphics[width = \linewidth]{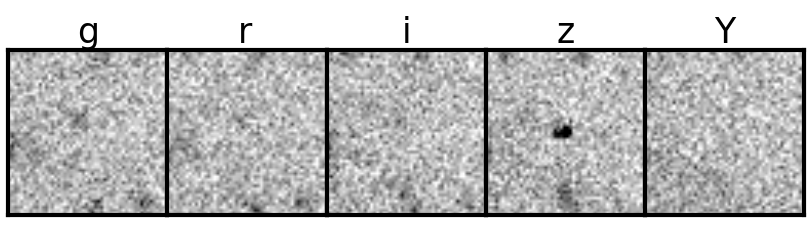}
\caption{This object has no discernible flux in the g band but a catalogue
entry of $g_{\mathrm{psf}} = 24.88 \pm 0.15$ and has no detection in the
r band. Despite appearing similar in both bands the object has been treated
differently by the pipeline in each band; leading to a detection in one and
not in the other. This demonstrates how the selection technique for dropouts
has to take into account the multiple ways of treating objects at the detection
threshold.} 
\label{Nogr}
\end{figure}

The cut of $\sigma_{g/r} > 0.1$ was used as it was empirically found from the
data that this was a good balance between keeping a few faint objects and not
throwing away objects that were falsely detected in g or r. Positive noise at
the site of the object can cause SExtractor to register a magnitude for the
object when there isn't a clean detection. This is shown in the cutouts in
figure \ref{Nogr}.

As high redshift quasars are unlikely to be detected in the g or r bands it is
important that objects with a false detection in g or r are kept.

\begin{figure}
\includegraphics[width = \linewidth]{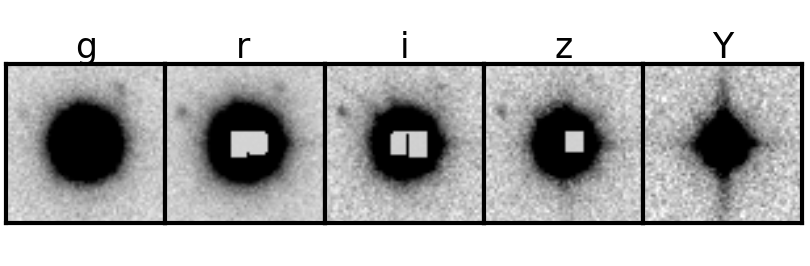}
\caption{Here an object which has saturated in the g and r bands is shown.
While it is obviously present in all the bands its PSF magnitudes are 99. This
objects has $g_{\mathrm{psf}} = 99.0 \pm 1.0$, $r_{\mathrm{psf}} = 99.0 \pm
1.0$, $i_{\mathrm{psf}} = 99.0 \pm 1.0$ and $z_{\mathrm{psf}} = 14.7326 \pm
0.0001$} \label{Brightgr}
\end{figure}

Another problem that was found was that some objects had a $g/r_{\mathrm{psf}}
= 99$ (which can mean a non detection) but were actually very bright, saturated
objects. An example of this is shown in figure \ref{Brightgr}.

The object in figure \ref{Brightgr} has the same r band magnitude as the object
shown in figure \ref{Nogr}. To distinguish between these two cases a rough
version of forced photometry was used as described in step 9 of section
\ref{SelectionCriteria}. As the difference in flux values for these two
categories of object is very dramatic and the r band magnitudes were not
required a complicated method of accurate forced photometry was not needed.

\subsection{Cosmic Ray Removal}

\begin{figure}
\includegraphics[width = \linewidth]{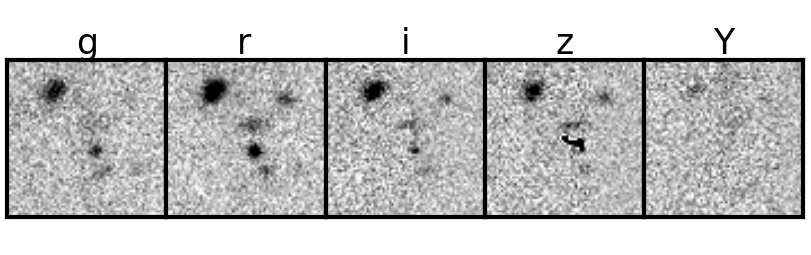}
\caption{Cosmic ray hits in the z band create objects that pass the selection
due to looking like extremely red i - z dropouts. These cutouts show the odd
shape of the cosmic ray as well as demonstrating that they are not smoothed out
by the PSF.} 
\label{CR}
\end{figure}

A cosmic ray hit in the z band can cause an object to pass the selection
criteria as it looks like a very extreme i - z dropout. An example cosmic ray
detection is shown in figure \ref{CR}, this shows the irregular shape of the
detection. 

The first step in removing these objects, which made up most of our sample
after step 7 in section \ref{SelectionCriteria}, was to require the objects to
have a Y band magnitude. As the Y band data is shallower and more variable than
the other bands no error cut was used. This left in objects that were false
detections in the Y band (such as the object in figure \ref{CR} which has
$Y_{\mathrm{psf}} = 20.54 \pm 1.0$) and the candidate list was still swamped
with cosmic rays. To remove these image based cosmic ray removal was used. This
step worked off the differences between adjacent pixels as it was found that
real objects had much shallower gradients between adjacent pixels than cosmic
rays.

\subsection{SQL Used}

Here we present the SQL code used to query the DES database. This
code returns 838 rows from a database with 40,129,963
unique sources as shown in Steps 1-8b in Table\ \ref{tab:selection}.

\begin{verbatim}
SELECT 

  * 

FROM 

  SVA1_COADD_GRIZY 

WHERE 

  /* Flag criteria */
  (FLAGS_G = 0 OR FLAGS_WEIGHT_G = 0 OR
   FLAGS_G > 3) AND 
  (FLAGS_R = 0 OR FLAGS_WEIGHT_R = 0 OR 
   FLAGS_R > 3) AND 
  (FLAGS_I = 0 OR FLAGS_WEIGHT_I = 0 OR 
   FLAGS_I > 3) AND 
  (FLAGS_Z = 0 OR FLAGS_WEIGHT_Z = 0 OR 
   FLAGS_Z > 3) AND 
  (FLAGS_Y = 0 OR FLAGS_WEIGHT_Y = 0 OR 
   FLAGS_Y > 3) AND 

  /* Z band limits */
  MAG_PSF_Z <= 21.5 AND MAGERR_PSF_Z < 0.1 
  AND 

  /* Morphological criterion */
  MAG_PSF_Z - MAG_MODEL_Z < 0.145 AND 

  /* I-Z colour criterion */
  MAG_PSF_I - MAG_PSF_Z > 1.694 AND

  /*  G and R dropout criteria */
  MAG_PSF_G > 23 AND MAG_PSF_R > 23 AND 
  MAGERR_PSF_G > 0.1 AND 
  MAGERR_PSF_R > 0.1 AND 

  /* Z-Y colour criterion */
  MAG_PSF_Z - MAG_PSF_Y < 0.5 AND

  /* Y band detection criterion */
  MAG_PSF_Y <> 99 AND 
  MAG_PSF_Y < 23

\end{verbatim}

\end{document}